\documentclass[aps,prl,twocolumn,superscriptaddress]{revtex4-1}
\usepackage{blindtext}
\usepackage{amsmath}
\usepackage{amssymb}
\usepackage{bm}
\usepackage{color}
\usepackage[normalem]{ulem} 
\usepackage{soul} 

\usepackage{graphicx}
\usepackage{indentfirst}
\usepackage{psfrag}
\usepackage{epsfig}

\begin{document}

\title{
		Isotropically polarized  speckle patterns 
}
\author{Mikolaj K. Schmidt}
\email{mikolaj\_schmidt@ehu.es}
\author{Javier Aizpurua}
\affiliation{Materials Physics Center CSIC-UPV/EHU, Paseo Manuel de Lardizabal 5, Donostia-San Sebastian, Spain}
\affiliation{Donostia International Physics Center DIPC, Paseo Manuel de Lardizabal 4, Donostia-San Sebastian, Spain}
\author{Xavier Zambrana-Puyalto}
\affiliation{Department of Physics and Astronomy, Macquarie University, North Ryde, New South Wales 2109, Australia}
\affiliation{ARC Center for Engineered Quantum Systems, Macquarie University, North Ryde, New South Wales 2109, Australia}
\author{Xavier Vidal}
\affiliation{Department of Physics and Astronomy, Macquarie University, North Ryde, New South Wales 2109, Australia}
\author{Gabriel Molina-Terriza}
\affiliation{Department of Physics and Astronomy, Macquarie University, North Ryde, New South Wales 2109, Australia}
\affiliation{ARC Center for Engineered Quantum Systems, Macquarie University, North Ryde, New South Wales 2109, Australia}
\author{Juan Jos\'e S\'{a}enz}
\email{juanjo.saenz@uam.es}
\affiliation{Donostia International Physics Center DIPC, Paseo Manuel de Lardizabal 4, Donostia-San Sebastian, Spain}
\affiliation{Depto. de F\'{i}sica de la Materia Condensada, Instituto Nicol\'{a}s Cabrera and Condensed Matter Physics Center (IFIMAC), Universidad Aut\'{o}noma de Madrid, 28049 Madrid, Spain}

\begin{abstract}
The polarization of the light scattered by an optically dense, random solution of dielectric nanoparticles shows peculiar properties when the scatterers exhibit strong electric and magnetic polarizabilities. While the distribution of the scattering intensity in these systems shows  the typical irregular \textit{speckle patterns}, the helicity of the incident light can be fully conserved when the electric and magnetic polarizabilities of the scatterers are equal. We show that the multiple scattering of helical beams by a random dispersion of ``dual'' dipolar nano-spheres leads to a speckle pattern exhibiting a perfect isotropic constant polarization, a situation that could be useful in coherent control of light as well as in lasing in random media.
\end{abstract}

\maketitle

The scattering of light by random media produces complex, irregular intensity distributions known as  {\em speckle patterns} \cite{dainty1984laser,goodman2007speckle}. Although the study of7 the statistical properties of  speckle patterns  has been a topic of high interest during the last decades, the statistics of the {\em polarization} of electromagnetic vector waves is still not well understood. Still, the depolarization of light in a random medium is the basis of an increasing  broad range  of applications from remote sensing \cite{carswell1980polarization}, enhanced backscattering phenomena  \cite{kuga1984retroreflectance,van1985observation,wolf1985weak,lenke2000multiple} or dynamic spectroscopy \cite{maret1987multiple,pine1988diffusing,weitz1993dynamic},   to biomedical imaging and diagnostics \cite{moscoso2001depolarization,angelsky2006polarization,pierangelo2011ex}. Even for static samples, the polarization of the scattered field is far from being isotropic \cite{carswell1980polarization} and the polarization of the speckle pattern may exhibit  rapid changes from one speckle grain to another \cite{zerrad2010gradual} with a nontrivial statistical distribution of polarization singularities \cite{schwartz2006backscattered,flossmann2008polarization}. 

It is generally assumed that multiple scattering of light from
inhomogeneities in optically dense media randomizes the
state of polarization of light. A wave propagating in such a
medium becomes rapidly depolarized in a characteristic length scale that  depends on  the properties of  both the scattering  medium and  the illuminating light \cite{bicout1994depolarization,gorodnichev1998diffusion,rojas2004depolarization}. 
Here we discuss a peculiar combination of random samples and laser beams that lead to  unusual ``anomalously'' polarized speckle patterns exhibiting  isotropic constant polarization.

Dielectric nanospheres of moderate permittivity like silicon \cite{evlyukhinoptical2010,garcia2011strong,schmidt2012dielectric} present strong magnetic and electric dipolar resonances in the visible, as well as in telecom and near-infrared frequencies, without spectral overlap between quadrupolar and higher-order modes. The interference between the electric and magnetic dipolar fields can lead to strongly asymmetric angular distributions of scattered intensity, including zero backscattering at specific wavelengths \cite{nieto-vesperinasangle-suppressed2011,gomez-medinaelectric2011,geffrinmagnetic2012,persondemonstration2013,fu2013directional} - the so-called, first Kerker's condition \cite{kerker1983electromagnetic}.  
In the dipolar approximation, such particles can be understood as ``dual'' scatterers, i.e. particles which are invariant under electromagnetic duality transformations. It should be noted that these scatterers are not ``dual" in terms of the ratio of their permittivity and permeability being equal to that of the surrounding medium \cite{fernandez-corbatonelectromagnetic2013,fernandez-corbatonhelicity2012}. Instead, the duality arises as a resonant effect due to electromagnetic modes of the scatterers \cite{zambrana-puyaltooe2013}. Then, the absence of backscattered light can be understood as a direct consequence of the simultaneous conservation of angular momentum and ``helicity'' in the scattering from cylindrically symmetric dual particles \cite{zambrana-puyaltoduality2013}. 

In this Letter, we study the light scattering on dielectric dual nanospheres illuminated by Bessel beams with a well-defined component of angular and linear momenta along the axis, and helicity \cite{fernandez-corbatonelectromagnetic2013,fernandez-corbatonhelicity2012}. Specifically, we analyze the far-field angular intensity distribution and polarization of the scattered light on a Si dipolar nanosphere as a function of its position from the axis. At the dual, first Kerker, condition, far-field light polarization  is independent on the scattering angle illustrating the conservation of the helicity in scattering by a single dual nanoparticle. Furthermore, we discuss the generalization of these results to systems comprising dimers \cite{albellalow-loss2013} and random ensembles of dual nanoparticles. Contrary to intuition,  we show that the multiple scattering of helical beams by a  random dispersion of dual dipolar nano-spheres leads to a speckle pattern exhibiting a perfect isotropic
  constant polarization.

\begin{figure}[htbp!]
\begin{center}
\includegraphics[width=\columnwidth]{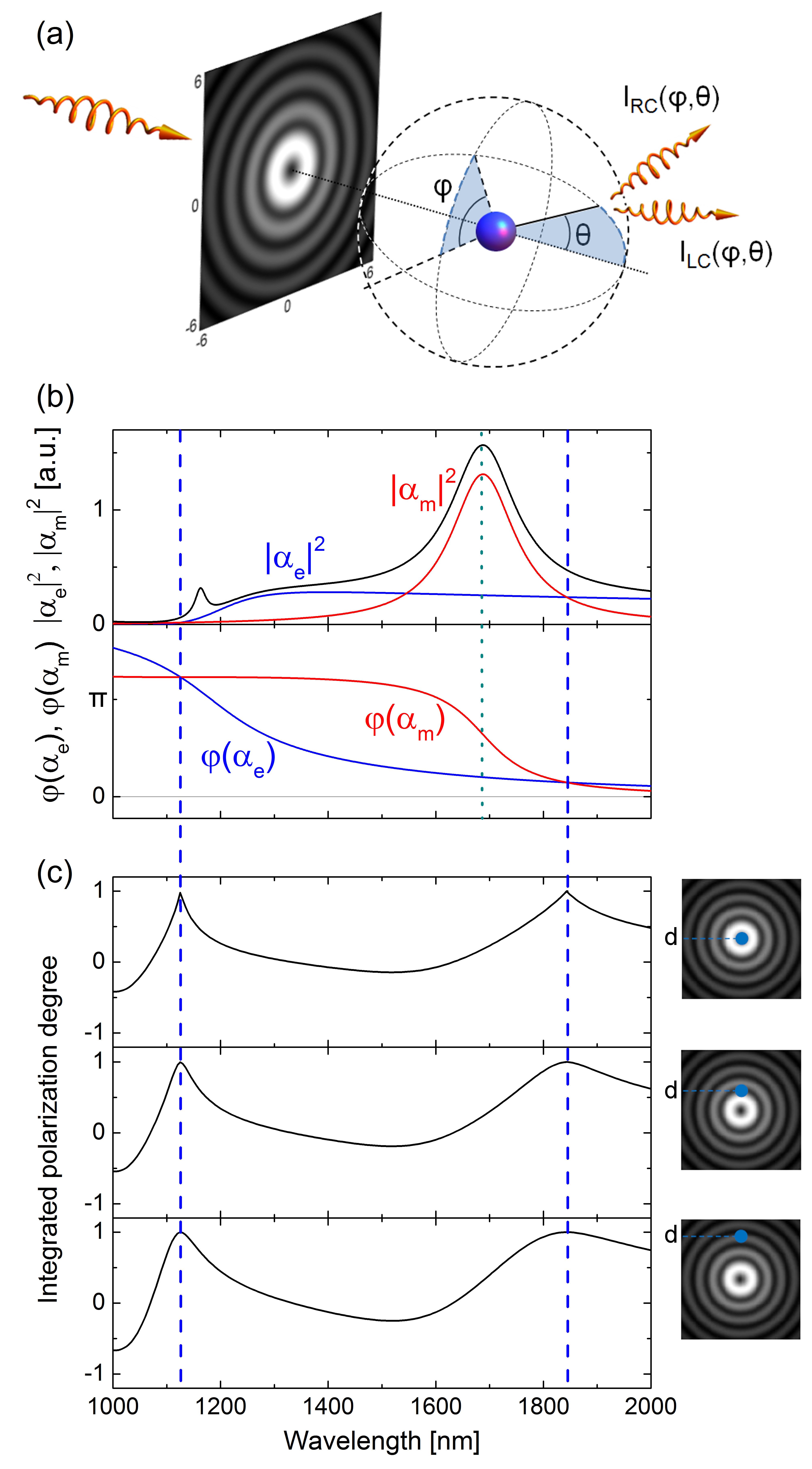}
\caption{(a) 
Schematic of the scattering process. 
(b) Amplitudes (top panel) and phases (bottom panel) of the electric ($\alpha_e$, blue line) and magnetic ($\alpha_m$, red line) polarizabilities of a 230 nm radius silicon sphere in the near-IR. Scattering cross section ($\sigma_{\text{scatt}} k^{-4}$ shown with black line) is dominated by the contributions from these dipolar terms. Dashed vertical lines indicate wavelengths at which the polarizabilities are equal both in magnitude and phase. (c) Integrated circular polarization factors $\eta_{tot}$ for the sphere positioned on the axis of the beam (top panel), and displaced by $2~\mu m$ (middle panel) and $4~\mu m$ (bottom panel) from the axis, as shown in the insets.
}
\end{center}
\end{figure}

Helical beams are a special class of solutions of the free-space Maxwell equations with the well-defined wavenumber $k$ and helicity $\Lambda$. The latter condition can be written down explicitly through the representation of the helicity operator, $\bm{\Lambda}$:
\begin{equation}\label{definition}
	\bm{\Lambda} \mathbf{E} = \frac{1}{k} \bm{\nabla} \times \mathbf{E} = \Lambda \mathbf{E},
\end{equation} 
with the allowed eigenvalues of $\Lambda=\pm 1$. One elementary solution to this set of equations takes the form of a circularly polarized planewave, for which the helicity can be identified with its handedness, giving $\Lambda=1(-1)$ for the left(right)-hand circularly polarized light. Furthermore, if the arbitrary helical beam is decomposed into a set of planewaves, all of them will exhibit circular polarization with identical handedness. 

In this work, when considering scattering into the far field, we will limit ourselves to investigating the handedness of the scattered field. This simplification stems from a simple observation:  any possible detector of scattered light will be measuring the properties of a single planewave, or a very narrow distribution of planewaves, scattered toward the detector's aperture.

However, before we consider scattering of helical beams, let us point out a very useful and crucial characteristic of such beams. Using the Faraday equation in frequency domain, we can relate the magnetic field of the beam with the curl, or - through the definition of the helicity operator - the helicity of the electric field:
\begin{equation}\label{eigenvalue} 
	\mathbf{H} = -\frac{i}{\mu_0 c} \mathbf{\Lambda} \mathbf{E}= -\frac{i}{\mu_0 c} \Lambda \mathbf{E},
\end{equation}
where $\mu$ is the vacuum permeability and $c$ - the speed of light in vacuum.

For illustrative purposes, throughout the manuscript we will be using a special form of the helical beams with an axial symmetry around the propagation direction $\hat{z}$, as defined in \cite{fernandez-corbatonhelicity2012}. We should stress that, although throughout this work we use a specific form of the incident helical light to illustrate the scattering processes, all of the results discussed in the paper are general and hold true for any given helical incident beam, for instance for Bessel beams with $n\neq 1$.

The intensity of the electric field of our helical beam, plotted in the schematic of Fig. 1(a), is invariant with respect to translation along $\hat{z}$. The spacing of the fringes in the intensity cross-section and its actual shape is determined by the helicity of the beam (here fixed as $\Lambda=-1$), its order ($n=1$), and by the aperture angle ($\theta_k=\pi/4$)~\cite{fernandez-corbatonhelicity2012}.
As we have mentioned above, in the scattering process, the helicity of light can be associated with the handedness of the scattered planewave. Therefore, to 
quantify the degree of helicity of the light scattered in the direction given by two angles $\varphi$ and $\theta$, we define a degree of helical polarization $\eta \in [-1,1]$:
\begin{equation} 
	\eta(\varphi,\theta) = \frac{I_{RC}(\varphi,\theta)-I_{LC}(\varphi,\theta)}{I_{RC}(\varphi,\theta)+I_{LC}(\varphi,\theta)},
\end{equation} 
where $I_{RC}$ and $I_{LC}$ are the polarization-resolved differential scattering cross sections for the right- and left-hand circularly polarized scattered light, respectively [see Fig. 1(a)]. Integrating $\eta$ over the two angles, we define the total degree of helical polarization $\eta_{tot} = (4\pi)^{-1}\int \int \sin(\theta) \eta(\varphi, \theta) d\theta d\varphi  \in [-1,1]$. We will use these quantities to determine and illustrate the conservation of helicity in scattering processes.

Having introduced the general properties of the incident beam, we now proceed to consider the so-called \textit{dual} scatterers. A particular example of a dual scatterer can be achieved with nanoparticles with electric and magnetic dipolar polarizabilities ($\alpha_e$ and $\alpha_m$) identical both in amplitude and in phase. In Fig.1(b) we plot the electric and magnetic polarizabilities of such a scatterer - the 230 nm radius silicon ($n=3.5$) sphere. Polarizabilities are equal for wavelengths of $\lambda=1844$ nm and $1160$ nm, marked with the vertical dashed blue lines. Note that for the latter case, the scattering spectrum is dominated by higher-order modes, and we cannot consider the spheres as dipolar scatterers. Furthermore, the spectra of $|\alpha_e|^2$ 
 and $|\alpha_m|^2$ cross also at $\lambda=1520$ nm, but for this wavelength the phases of the polarizabilities are different (i.e. $\text{Re}\{\alpha_e\}=-\text{Re}\{\alpha_m\}$, which corresponds to the almost-zero-forward condition \cite{nieto-vesperinasangle-suppressed2011,gomez-medinaelectric2011,geffrinmagnetic2012}). Polarizabilities and the scattering cross section plotted in Fig.1(b) have been obtained from the Mie theory.

In the spectral range where the scattered fields can be described by dipolar electric and magnetic responses, the polarization-resolved differential cross section takes the following, analytical form \cite{jackson1962classical}
\begin{equation} 
	I_{\mathbf{\epsilon}}(\mathbf{n}) \propto |\bm{\epsilon}^* \cdot \mathbf{p} + (\mathbf{n} \times \bm{\epsilon}^*) \cdot \mathbf{m}/c |^2,
\end{equation} 
with $\bm{\epsilon}$ corresponding to the polarization of the scattered light, and $\mathbf{p}$ and $\mathbf{m}$ denoting the electric and magnetic dipoles, respectively. The dipoles are induced by the incident fields ($\mathbf{E}$, $\mathbf{H}$):
\begin{equation} \label{dipoles}
	\mathbf{p} = \varepsilon_0 \alpha_e \mathbf{E} \quad , \quad  \mathbf{m} = \alpha_m \mathbf{H}.
\end{equation} 
For the dual nanoparticle ($\alpha_e=\alpha_m=\alpha_0$), using the relationship Eq. (\ref{eigenvalue}), we get:
\begin{equation} \label{cs}
	I_{\bm{\epsilon}}(\mathbf{n}) \propto \left| \varepsilon_0 \alpha_0^* \mathbf{E}^* \cdot \left[ \bm{\epsilon} + i \Lambda \  \mathbf{n} \times \bm{\epsilon} \right]  \right|^2.
\end{equation} 
We want to investigate the polarization of the scattered light in the basis of right-hand circularly polarized $\mathbf{r}$ (RC) and left-hand circularly polarized $\mathbf{l}$ (LC) light. For any scattering direction $\mathbf{n}$, we have $\mathbf{n} \times \mathbf{l} = -i \mathbf{l}$ and $\mathbf{n} \times \mathbf{r} = i \mathbf{r}$. Thus, the squared expression in the scattering cross section (Eq. (\ref{cs})) is proportional to
\begin{equation} 
	{{I_{LC}}\choose{I_{RC}}} \propto \left| {{\mathbf{l}}\choose{\mathbf{r}}} + \Lambda {{\mathbf{l}}\choose{-\mathbf{r}}}\right|^2,
\end{equation} 
indicating that for the $\Lambda=1(-1)$ incident beams the RC (LC) polarization of the scattered light vanishes. Therefore, for the dual nanoparticle and the incident helical field, $\eta(\mathbf{n})$ for every direction should be equal to 1, giving $\eta_{tot}=1(-1)$ for the incident $\Lambda=-1(1)$ light.

We illustrate this scattering invariant in Fig. 1(c). In the plots we present the spectra of the total degree of helical polarization $\eta_{tot}$, calculated for the scatterer positioned on the axis of the beam (top panel), or shifted away from it by $d=2~\mu$m (middle panel) or $d=4~\mu$m (bottom panel), as shown schematically in the insets. For the two wavelengths at which the two polarizabilities match ($\lambda = 1160$ nm and $1844$ nm), $\eta_{tot}$ reaches its maximum value $1$, indicating a fully circular polarization of the scattered light. 

The scattering of a dual nanoparticle preserves the helicity everywhere and not only the far field components as we have shown. As the scattered near-field, composed primarily of the evanescent waves is essential for the understanding of systems comprised of many scatterers, below we will investigate in detail the helicity of the entirety of the scattered field for the dipolar spheres.

To arrive at this result, we consider the relationship between the electric $\mathbf{p}$ and magnetic $\mathbf{m}$ dipoles induced in a dual nanoparticle by helical light. Inserting Eq.(\ref{eigenvalue}) into definitions given in Eq.(\ref{dipoles}), we arrive at
\begin{equation} 
\label{dualdipole}
	\mathbf{m} = - i c \Lambda \mathbf{p}.
\end{equation} 
The scattered electric field from such a pair of dipoles can be expressed through Green's functions as
\begin{equation} 
	\mathbf{E}_{scatt} = \frac{k^2}{\varepsilon_0} \mathbf{G}_{E}  \mathbf{p} + i Z k^2 \mathbf{G}_M  \mathbf{m} = \frac{k^2}{\varepsilon_0} (\mathbf{G}_{E}  + \Lambda \ \mathbf{G}_M ) \mathbf{p}.
\end{equation} 
To calculate the action of the helicity operator on $\mathbf{E}_{scatt}$, we use the following property of the Green's function;
\begin{equation} 
	\mathbf{\Lambda} \mathbf{G_E} = \mathbf{G_M} \quad , \quad \mathbf{\Lambda} \mathbf{G_M} = \mathbf{G_E},
\end{equation} 
which can be derived taking the definition of the helicity operator, Eq.(1). We then have
\begin{equation} 
	\mathbf{\Lambda} \mathbf{E}_{scatt} = \frac{k^2}{\varepsilon_0} (\mathbf{G}_{M}  + \Lambda \mathbf{G}_E ) \mathbf{p}.
\end{equation} 
Since the eigenvalues of the helicity operator follow $\Lambda^2=1$, we can rewrite the above equations as
\begin{equation} 
	\mathbf{\Lambda} \mathbf{E}_{scatt} = \Lambda \frac{k^2}{\varepsilon_0} (\Lambda \mathbf{G}_{M}  + \mathbf{G}_E ) \mathbf{p} = \Lambda  \mathbf{E}_{scatt}.
\end{equation} 
This result is not so surprising if we consider that Eq. (\ref{dualdipole}) represent the two only kind of dual dipoles \cite{PhysRevB.2013.dualdip}. Then, when the dipolar moments dominate, Eq. (\ref{dualdipole}) ensures that the helicity is preserved everywhere (near and far field).

\begin{figure}[htbp]
\centering
\includegraphics[width=\columnwidth]{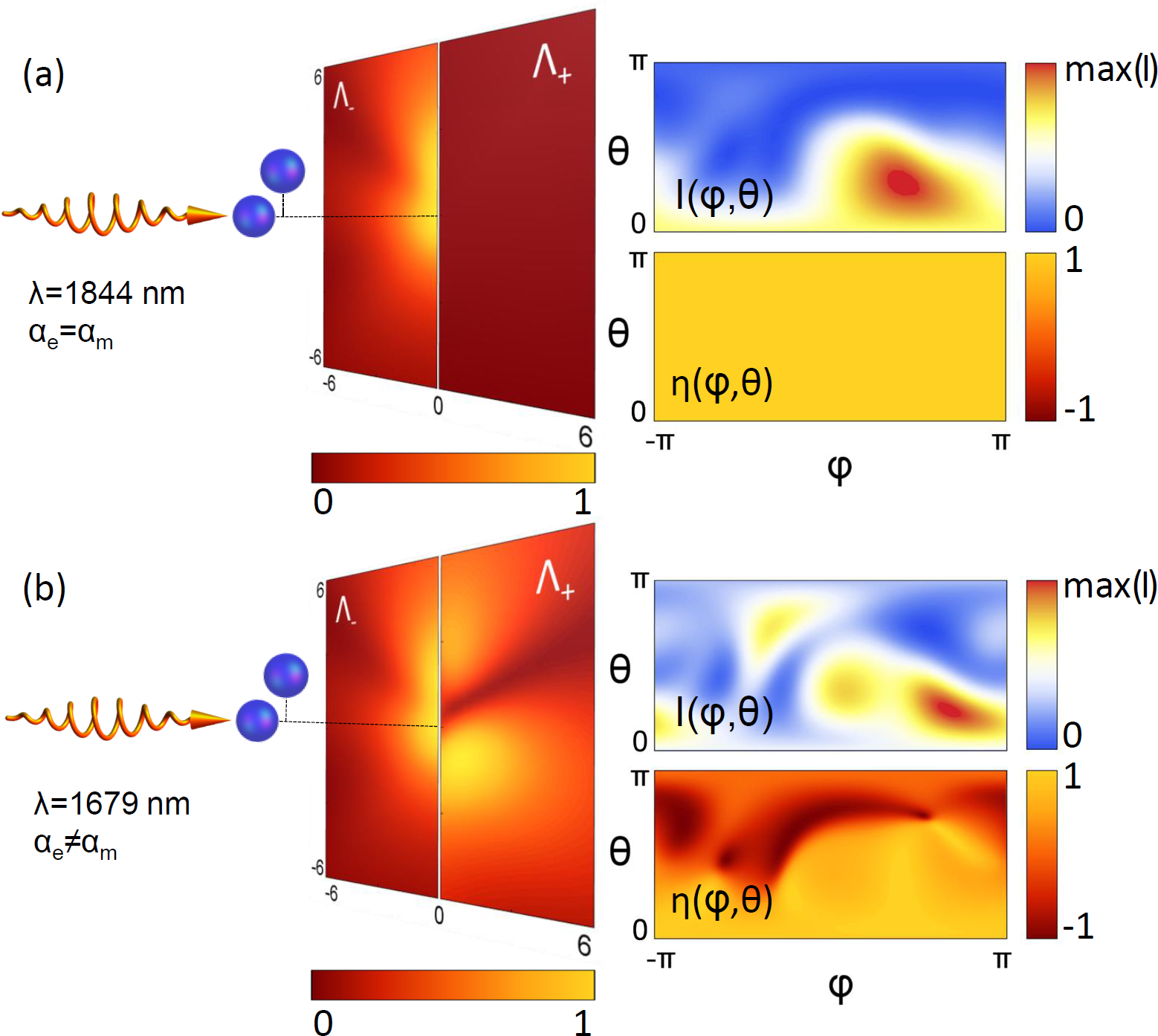}
\caption{Light scattering by a dimer of spheres. The incident helical ($\Lambda=-1$) light of (a) 1844 nm or (b) 1679 nm wavelength is scattered on a dimer of two (a) dual and (b) non-dual silicon spheres. One of the spheres is positioned at the origin of the coordinates system and the other on is shifted from it  $0.5~\mu$m along the axis and $1~\mu$m in the transverse directions. In the near field (calculated at $z=3~\mu$m) the helicities of the scattered light can be measured by calculating distributions $\Lambda_\pm$ (central panels). Vanishing distribution of $\Lambda_+$ in (a)  indicates that the scattered light retains the negative helicity of the incident light, while the non-vanishing distributions in (b) indicate mixing of the $\Lambda=-1$ and $\Lambda=1$ components of light. Right panels represent the differential scattering cross sections of the scattered light $I(\varphi,\theta)$ (upper plots) and of the polarization degree $\eta(\varphi,\theta)$ (lower panel). For the dual spheres (a) $\eta(\varphi,\theta) = 1$ is a constant function,  indicating that the scattered light is fully circularly polarized ($\eta_{tot}=1$).}
\end{figure}

Since the helicity of the electromagnetic field is conserved in the process of scattering on a single dual scatterer, it should also be conserved in the subsequent scattering events on other dual scatterers. To illustrate this helicity invariance, we consider the scattering of the helical beam on a dimer of the modeled silicon spheres \cite{albellalow-loss2013}. The central panels in Fig. 2 show the distribution of the two quantities defined as
\begin{equation} 
	\Lambda_\pm = \left|\mathbf{E}_{scatt} \pm i c \mu_0 \mathbf{H}_{scatt} \right|^2,
\end{equation} 
calculated in the transverse plane defined by $z=3~\mu$m. 
 $\Lambda_\pm$ computes the scattered intensity into modes with helicity $\pm$. 
Two wavelengths of the incident light are considered: (a) $\lambda=1844$ nm, for which the scatterers are dual, and (b) $\lambda=1679$ nm, at which the scatterers have a dominating magnetic dipolar response. For a helical beam with $\Lambda=-1$ interacting with a dual dimmer,   the scattered intensity into modes with $\Lambda_+$ will be zero as shown in Fig. 2(a). 
 In contrast, for the non-dual scatterers (Fig. 2(b)), neither one of the two fields $\Lambda_\pm$ vanishes, indicating the mixing of the two helicities in the scattering process.
For both wavelengths, we also investigate the far-field properties of the scattered light, plotting its differential scattering cross section $I(\varphi,\theta)$ and the degree of helical polarization $\eta(\varphi,\theta)$ in the right panels of Fig. 2. The scattered light is shown to be fully circularly polarized only for the dual sphere (Fig.2(a)). 

We can also extend our considerations to the random media, modeled as a distribution of the dual scatterers, where each one preserves the helicity in a single scattering event. By using a coupled electric and magnetic dipole method  \cite{markel1993coupled,mulholland1994light,chaumet2000coupled}, we illustrate this process in Fig. 3 where we investigate the scattering of incident helical light on a random distribution of 80 nanoparticles positioned randomly in a cubic volume of 60~$\mu$m edge length.   Similarly as in Fig. 2, we consider two wavelengths of incident light: (a) 1844 nm, at which the scatterers are dual, and (b) 1679 nm. In the former case, the circular polarization degree $\eta$ is constant and equal to 1 for any scattering direction (right bottom panel in (a)), indicating the conservation of helicity in the multiscattering process. For the non-dual scatterers, polarization degree does not exhibit any significant preservation of helicity. 
\begin{figure}[htbp]
\centering
\includegraphics[width=\columnwidth]{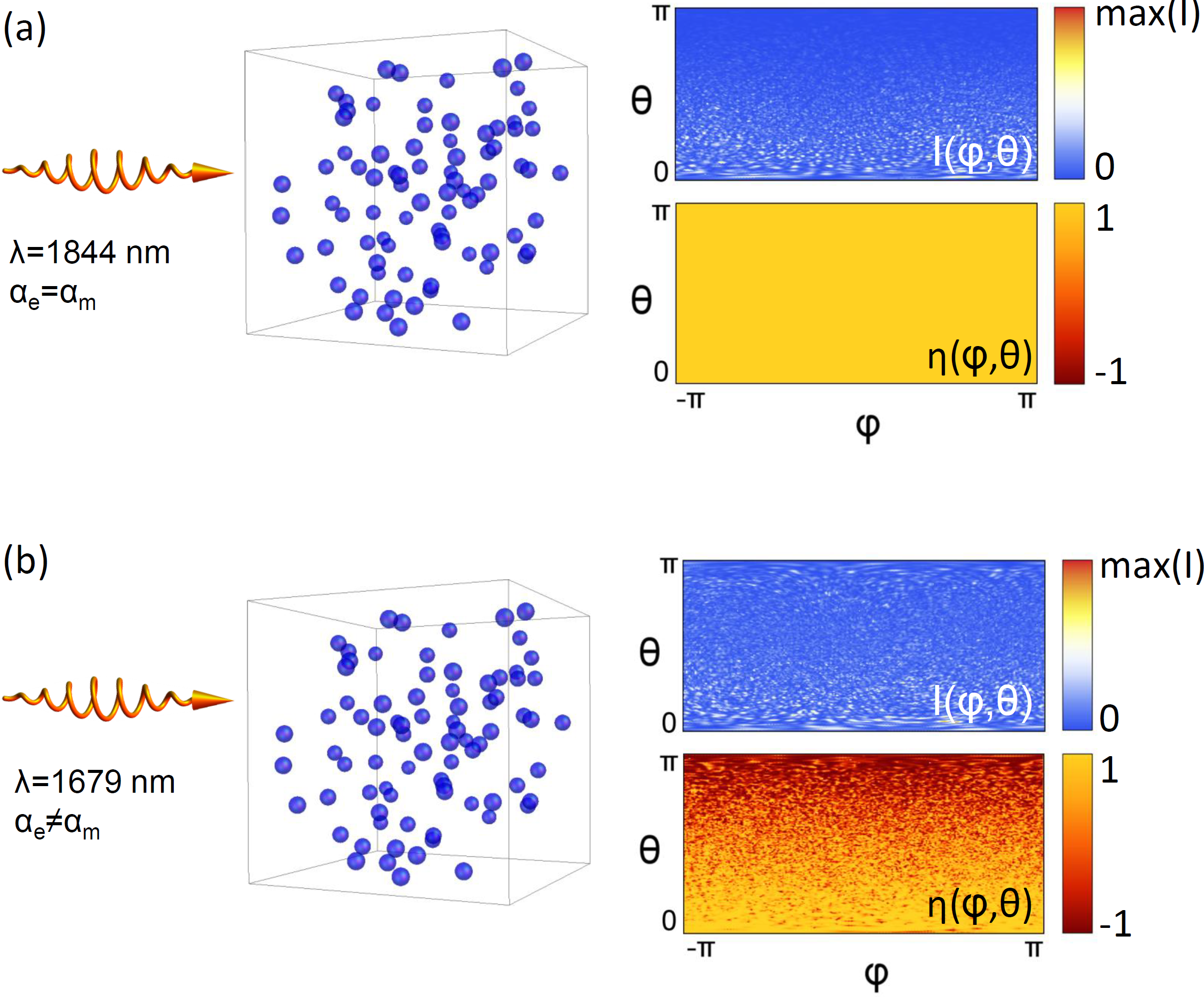}
\caption{Light scattering by a random medium. The incident helical ($\Lambda=-1$) light of (a) 1844 nm or (b) 1679 nm wavelength scatters on an ensemble of 80 randomly distributed (a) dual and (b) non-dual silicon spheres. The scatterers were randomly distributed in a cubic box with edges of 60 $\mu$m length, centered on the axis of the beam. Panels on the right represent the differential scattering cross-section $I(\varphi,\theta)$ (upper plots) and of the polarization degree $\eta(\varphi,\theta)$. }
\end{figure}

It is worth noticing that the intensity distribution for dual particles presents a clear asymmetry between forward and backward scattering. Due to the conservation of angular momentum and helicity, a complete suppression of backscattering (at $\theta=0$) is expected \cite{fernandez-corbatonelectromagnetic2013} for dual and axially symmetric samples. The partial (statistically averaged) axial symmetry particle distribution explains the observed results. Such an asymmetry is not observed for the non-dual medium (Fig. 3(b)).

Finally, we briefly discuss the robustness of the helicity conservation against the deviation from the perfect duality condition of the scattering medium. Such deviation naturally arises when nanoparticles in the solution are not identical, but represent a distribution of radii with some standard deviation $\sigma$. We have carried out simulations of random media comprising scatterers with radii distributed around $r_0=0.23 \mu m$ with $\sigma=0.02 r_0$ and $0.05 r_0$. The integrated helicity degree was in result reduced to $0.85$ or $0.6$, respectively. A complete set of results and more elaborate discussion is included in the Supplemental Material \cite{SM}.

In conclusion, we have investigated the problem of scattering of helical light by dielectric nanoparticles exhibiting strong electric and magnetic activity. At the Kerker condition, when both electric and magnetic polarizabilities are equal, the scattering preserves the helicity and polarization of light. 
We have shown this anomalous conservation of the scattering polarization in the case of a single nanoparticle, a dimer and a random solution of dielectric nanoparticles. Our results open a pathway to exploit novel properties in random scattering media, including intriguing applications in random lasing \cite{wiersma2008physics,gottardo2008resonance}, as well as provide new possibilities to characterize magnetic optical properties of nanoscatterers \cite{zia11}. 

MKS and JA acknowledge funding from the project FIS2013-41184-P of the Spanish Ministry of Economy and Competitiveness, the ETORTEK IE14-393 NANOGUNE'14 project of the Department of Industry of the Government of the Basque Country, project IT756-13 of the Department of Education and Culture of the Basque Country, and scholarship AP-2012-4204 from the Spanish Ministry of Education, Culture and Sport. JJS acknowledges financial support by the Spanish MINECO  (Grant No. FIS2012-36113) and by  an IKERBASQUE Visiting Fellowship. GMT is an Australian Research Council Future Fellow.

\bibliographystyle{apsrev4-1} 
\bibliography{helicity_bib13}

\end{document}